\documentclass[times, 10pt,twocolumn]{article} 
\usepackage{times}
\usepackage{epsfig}

\newcommand{\ie}{{\it i.e.,$\ $}}
\newcommand{\eg}{{\it e.g.,$\ $}}

\begin{document}

\thispagestyle{empty}
\setcounter{page}{1}
\twocolumn
\normalsize

\title{Integrity-Enhancing Replica Coordination for Byzantine Fault Tolerant Systems\thanks{This work is sponsored by Cleveland State University through a Faculty 
Research Development award.}}

\author{Wenbing Zhao\\
Department of Electrical and Computer Engineering\\ 
Cleveland State University\\
2121 Euclid Ave, Cleveland, OH 44115\\
wenbing@ieee.org\\
}

\maketitle
\thispagestyle{empty}

\begin{abstract}

Strong replica consistency is often achieved by writing deterministic 
applications, or by using a variety of mechanisms to render replicas 
deterministic. There exists a large body of work on how to render replicas 
deterministic under the benign fault model. However, when replicas can 
be subject to malicious faults, most of the previous work is no longer 
effective. Furthermore, the determinism of the replicas is often considered 
harmful from the security perspective and for many applications, their 
integrity strongly depends on the randomness of some of their internal 
operations. This calls for new approaches towards achieving replica 
consistency while preserving the replica randomness. In this paper, 
we present two such approaches. One is based on Byzantine agreement and the 
other on threshold coin-tossing. Each approach has its strength and 
weaknesses. We compare the performance of the two approaches and outline 
their respective best use scenarios.
\end{abstract}


\section{Introduction}

Strong replica consistency is an essential property for replication-based 
fault tolerant distributed systems. It can be achieved via a number of 
different techniques. In this paper, we investigate the challenges in 
achieving integrity-preserving strong replica consistency and present our 
solutions for state-machine based Byzantine fault tolerant 
systems~\cite{bft-acm}. While it is widely known that strong replica 
consistency can also be achieved through the systematic-checkpointing
technique~\cite{delta4} for nondeterministic applications in the benign
fault model, it is generally regarded as too expensive and it is not
suitable for Byzantine fault tolerance.

The state-machine based approach is one of the fundamental techniques 
in building fault tolerant systems~\cite{statemachine}. In this approach, 
replicas are assumed to be either deterministic or rendered-deterministic. 
There has been a large body of work on how to render replicas deterministic 
in the presence of replica nondeterminism under the benign fault 
model (\eg ~\cite{bft-acm,base,delta4,priya06}). 
However, when the replicas can 
be subject to Byzantine faults, which is the case for many Internet-based 
systems, most of the previous work is no longer effective. Furthermore, the 
determinism (or rendered-determinism) of the replicas is often considered 
harmful from the security perspective (\eg with replication, an adversary
can compromise any of the replicas to obtain confidential 
information~\cite{deswarte}) and for many applications, their 
integrity is strongly dependent on the randomness of some of their internal 
operations (\eg random numbers are used for unique identifier generation 
in transactional systems and for shuffling cards in online poker games, 
and if the randomness
is taken away by a deterministic algorithm to ensure replica consistency, 
the identifiers or the hands of cards can be made predictable,
which can easily lead to exploit~\cite{ssbook,cryptovirology}).
This calls for new approaches towards achieving strong replica 
consistency while preserving the randomness of each replica's operations. 

In this paper, we present two alternative approaches towards our goal. 
The first one is based on Byzantine agreement ~\cite{bft-acm} 
(referred to as the BA-algorithm in this paper) and the other on a 
threshold coin-tossing scheme~\cite{ctba} (referred to as the CT-algorithm). 
Both approaches rely on a collective determination for decisions involving 
randomness, and the determination is based on the contributions made by 
a set of replicas (at least one of which must be correct), to 
avoid the problems mentioned above. They differ mainly by how the 
collective determination is carried out. In the BA-algorithm,
the replicas first reach a Byzantine agreement on the set of contributions
from replicas, and then apply a deterministic algorithm (for all practical 
purposes, the bitwise exclusive-or operation~\cite{cryptovirology}) 
to compute the final random value.
The CT-algorithm uses the threshold coin-tossing scheme introduced
in~\cite{ctba} to derive the final random value, without the need
of a Byzantine agreement step. Even though the CT-algorithm saves on
communication cost, it does incur significant computation overhead due to
the CPU-intensive exponentiation calculations. Consequently, as we will show
in Section~\ref{implsec}, the BA-algorithm performs the best in a 
Local-Area Network
(LAN) environment, where the CT-algorithm is more appropriate for the
Wide-Area Network (WAN) environment where message passing is expensive. 
Furthermore, to ensure the freshness
of the random numbers generated, the replicas using the BA-algorithm
should have access to high entropy sources (which is relatively easy
to satisfy) and the replicas should be able to refresh their key shares
periodically in the CT-algorithm. For the latter, we envisage that
a proactive threshold signature scheme could be 
used~\cite{threshsig.b,threshsig.f,threshsig.g}. However,
the discussion of proactive threshold signature techniques is out of the
scope of this paper. 

To summarize, we make the following research contributions in this paper: 
\begin{itemize}
\item We point out the danger and pitfalls of controlling replica randomness
for the purpose of ensuring replica consistency. Removing randomness 
from replica operations (when it is needed) could seriously compromise the 
system integrity.

\item We propose the use of collective determination of random numbers
contributed from replicas, as a practical
way to reconcile the requirement of strong replica consistency and
the preservation of replica randomness.

\item We present a light-weight, Byzantine agreement based algorithm to
carry out the collective determination. The BA-algorithm only introduces
two additional communication steps because the Byzantine agreement for
the collective determination of random numbers can be integrated into
that for message total ordering, as needed by the state-machine replication.
The BA-algorithm is particularly suited for Byzantine fault tolerant systems
operating in the LAN environment, or where replicas 
are connected by high-speed low-latency networks.

\item We further present an algorithm that uses the threshold coin-tossing 
scheme~\cite{ctba} as an alternative method for collective
determination of random numbers. The coin-tossing scheme is
introduced in~\cite{ctba} as an instrumental mechanism for a group of 
replicas
to reach Byzantine agreement in asynchronous systems. To the best of our
knowledge, our work is the first to show its usefulness in helping 
to ensure strong replica consistency without compromising the system
integrity.

\item We conduct extensive experiments, in both a LAN testbed and
an emulated WAN environment, to thoroughly characterize the 
performance of the two approaches. 
\end{itemize}

\section{Byzantine Fault Tolerance}

In this section, we introduce the system model for our work, and the
practical Byzantine fault tolerance algorithm (BFT algorithm, for short) 
developed by Castro and
Liskov~\cite{bft-acm} as necessary background information.

Byzantine fault tolerance refers to the capability of a system to tolerate
Byzantine faults. It can be achieved by replicating the server and
by ensuring that all server replicas reach an agreement on the total ordering
of clients' requests despite the existence of Byzantine faulty replicas 
and clients. Such an agreement is often referred to as Byzantine 
agreement~\cite{lamport:byz}.

In recent several years, a number of efficient Byzantine agreement 
algorithms~\cite{bft-acm,speculativebft,alvisi-bft} have been proposed. 
In this work, we focus on the BFT algorithm
and use the same system model as that in~\cite{bft-acm}.

The BFT algorithm operates in an asynchronous distributed environment.
The safety property of the algorithm, \ie all correct replicas agree
on the total ordering of requests, is ensured without any
assumption of synchrony. However, to guarantee liveness, \ie for the
algorithm to make progress towards the Byzantine agreement, certain
synchrony is needed. Basically, it is assumed that the message transmission 
and processing delay has an asymptotic upper bound. This bound is 
dynamically explored in the algorithm in that 
each time a view change occurs, the timeout for the new view is doubled.

The BFT algorithm is executed by a set of $3f+1$ replicas to tolerate up 
to $f$ Byzantine faulty replicas. One of the replicas is designated as 
the primary
while the rest are backups. Each replica is assigned a unique id $i$, 
where $i$ varies from $0$ to $3f$. For view $v$, the replica whose id $i$ 
satisfies $i=v \: mod \: (3f+1)$ would serve as the primary. The view 
starts from 0. For each view change, the view number is increased by one 
and a new primary is selected.

The normal operation of the BFT algorithm involves three phases. During the 
pre-prepare phase, the primary multicasts 
a pre-prepare message containing the client's request, the current view and 
a sequence number assigned to the request to all backups. A backup verifies 
the request and the ordering information. If the backup accepts the 
pre-prepare message, it multicasts a prepare message containing the ordering
information and the digest of the request being ordered.
This starts the prepare phase. A replica waits until
it has collected $2f$ matching prepare messages from different replicas,
and the pre-prepare message, before it multicasts a commit message to other 
replicas, which starts the commit phase. The commit phase 
ends when a replica has collected $2f+1$ matching commit messages from 
different replicas (possibly including the one sent or would have been sent
by itself). At this point, the request message has been totally ordered and 
it is ready to be delivered to the server application once all previous
requests have been delivered.

All messages exchanged among the replicas, and those between the replicas 
and the clients are protected by an authenticator~\cite{bft-acm} (for
multicast messages), or by a message authentication code (MAC)
(for point-to-point communications). An authenticator is formed by a number
of MACs, one for each target of the multicast.
We assume that the replicas and the
clients each has a public/private key pair, and the public keys are known
to everyone. These keys are used to generate symmetric keys needed
to produce/verify authenticators and MACs. To ensure freshness, the 
symmetric keys are periodically refreshed by the mechanism described 
in~\cite{bft-acm}. We assume that the adversaries have limited computing 
power so that they cannot break the security mechanisms described above.

Furthermore, we assume that a faulty replica cannot transmit
the confidential state, such as the random numbers collectively determined, 
to its colluding clients in real time.
This can be achieved by using an application-level gateway,
or a privacy firewall as described by Yin et al.\cite{alvisi-bft}, to
filter out illegal replies. A compromised replica may, however,
replace a high entropy source to which it retrieves random numbers
with a deterministic algorithm, and convey such an algorithm via 
out-of-band or covert channels to its colluding clients.

\section{Pitfalls in Controlling Replica Randomness}

In this section, we analyze a few well-known approaches possibly be
used to ensure replica consistency in the presence of replica 
randomness. We show that they are not robust against Byzantine faulty 
replicas and clients.

For replicas that use a pseudo-random number generator, they can be
easily rendered deterministic by ensuring that they use the same seed
value to initialize the generator. One might attempt to use the sequence
number assigned to the request as the seed. Even though this approach
is perhaps the most economical way to render replicas deterministic
(since no extra communication step is needed and no extra information
is to be included in the control messages for total ordering of requests),
it virtually takes the randomness away from the fault tolerant systems.
In the presence of Byzantine clients, the vulnerability can be exploited
to compromise the integrity of the system. For example, a Byzantine faulty
client in an online
poker game can simply try out different integer values as the seed to the 
pseudo-random generator (if it is known to the client) to guess the
hands of the cards in the dealer and compare with the ones it has gotten.
The client can then place its bets accordingly and gain unfair
advantage.

A seemingly more robust approach is to use the timestamp as the seed
to the pseudo-random number generator. As shown 
in~\cite{ssbook,cryptovirology}, the use of
timestamp does not offer more robustness to the system because
it can also be guessed by Byzantine faulty clients. Furthermore, the
use of timestamp imposes serious challenges in asynchronous distributed
systems because of the requirement that all replicas must use the same
timestamp to seed the pseudo-random number generator. In~\cite{bft-acm},
a mechanism is proposed to handle this problem by asking the primary to
piggyback its timestamp, to be used by backups as well, with the 
pre-prepare message. However, the issue is that the backups
have very limited ways of verifying the timestamp proposed (other than that
the timestamp must be monotonically increasing) without resorting to
strong synchrony assumptions (such as bounds on processing and
message passing).

The only option remaining seems to be the use of a truly random number to 
seed the pseudo-random number generator (or to obtain random numbers entirely
from a high entropy source). We note that the elegant mechanism described 
in~\cite{bft-acm} cannot be used in this case 
because backups have no means to 
verify whether the number proposed by the primary is taken from a 
high-entropy source, or is generated according to a deterministic algorithm. 
If the latter is the case, the Byzantine faulty primary could continue 
colluding with Byzantine faulty clients without being detected.

Therefore, we believe the most effective way in countering such threats is
to collectively determine the random number, based on the contributions
from a set of replicas so that Byzantine faulty replicas cannot influence
the final outcome. The set size depends on the algorithms used, as we will
show in the next two sections, but it must be greater than the number
of faulty replicas tolerated ($f$) by the system.

\begin{figure}[t]
  \center
   \includegraphics[width=3.0in]{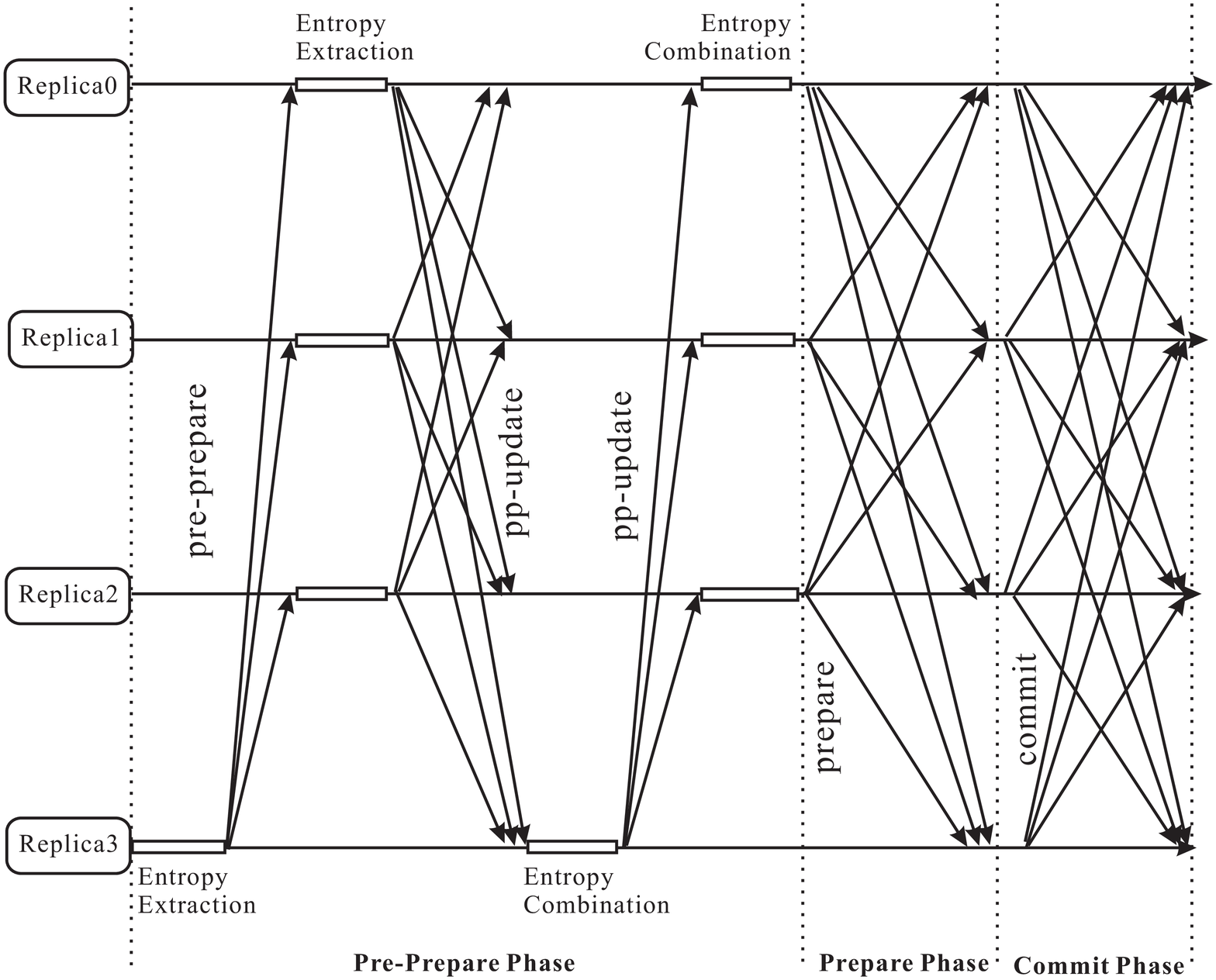}
   \caption{Normal operation of the BA-algorithm.}
   \label{bafig}

\end{figure}

\section{The BA-Algorithm}

The normal operation of the BA-algorithm is illustrated in 
Figure~\ref{bafig}. As can be seen, the collective-determination mechanism 
is seamlessly integrated into the original BFT algorithm. 
On ordering a request, the primary determines the order of the request 
(\ie assigns a sequence number to the request), and queries
the application for the type of operation associated with the request.
If the operation involves with a random number as input, the primary
activates the mechanism for the BA-algorithm. The primary then obtains its 
share of random number by extracting from its own entropy source, and 
piggybacks the share with the pre-prepare message multicast to all backups. 
The pre-prepare message has the form 
$<${\sc pre-prepare},$v,n,d,R_p$$>$$\alpha_{p}$, where $v$ is the view 
number, $n$ is the sequence number assigned to the request, $d$ is
the digest of the request, $R_p$ is the random number generated by the
primary, and $\alpha_{p}$ is the authenticator for the message.

On receiving the pre-prepare message, a backup performs the usual chores
such as the verification of the authenticator before it
accepts the message. It also checks if the request will indeed trigger
a randomized operation, to prevent a faulty primary from putting
unnecessary loads on correct replicas (which could lead to a denial of 
service attack). If the pre-prepare message is acceptable, the replica
creates a pre-prepare certificate for storing the relevant information, 
generates a share of random number from its entropy source, and multicasts 
to all replicas a pp-update message, in the form
$<${\sc pp-update,}$v,n,i,R_i,d$$>$$\alpha{_i}$, where $i$ is the sending
replica identifier, $R_i$ is the random number contributed by replica $i$.

When the primary has collected $2f$ pp-update messages, it combines
the random numbers received according to a deterministic algorithm
(referred to as the entropy combination step
in Figure~\ref{bafig}), and builds a
pp-update message with slightly different content than those sent by
backups. In the pp-update message sent by the primary, the $R_i$ component
is replaced by a set of $2f+1$ tuples containing the random numbers 
contributed by replicas (possibly including its own share), $S_R$. 
Each tuple has the form
$<$$R_i,i$$>$. The replica identifier is included in the tuple to
ease the verification of the set at backups. 

On receiving a pp-update message, a backup accepts the message and stores
the message in its data structure provided that the message has a correct
authenticator, it is in view $v$ and it has accepted a pre-prepare message
to order the request with the digest $d$ and sequence number $n$. A backup
proceeds to the entropy combination step only if (1) it has accepted a
pp-update message from the primary, and (2) $2f$ pp-update messages
sent by the replicas referenced in the set $S_R$. The backup requests
a retransmission from the primary for any missing pp-update message.

After the entropy combination step is completed, a backup multicasts a 
prepare message in the form $<${\sc prepare}$v,n,i,d'$$>$$\alpha_{i}$, 
where $d'$ is the digest of the request concatenated by the combined random 
number.

When a replica has completed the entropy combination step, and it has
collected $2f$ valid prepare messages from different replicas (possibly
including the message sent or would have been sent by itself), 
it multicasts to
all replicas a commit message in the form 
$<${\sc commit}$v,n,i,d'$$>$$\alpha_{i}$. When a replica receives $2f+1$
valid commit messages, it decides on the sequence number
and the collectively determined random number. At the time of
delivery to the application, both the request and the random number
are passed to the application.

In Figure~\ref{bafig}, the duration of the entropy extraction and 
combination 
steps have been intentionally exaggerated for clarify. In practice, the
entropy combination can be achieved by applying a bitwise exclusive-or
operation on the set of random numbers collected, which is very fast. The
cost of entropy extraction depends on the scheme used. Some schemes, such
as the TrueRand method~\cite{truerand}, allows very prompt entropy 
extraction. TrueRand works by gathering the underlying randomness from 
a computer by measuring the drift between the system clock and the 
interrupts-generation rate on the processor.

\begin{figure}[t]
  \center
   \includegraphics[width=2.6in]{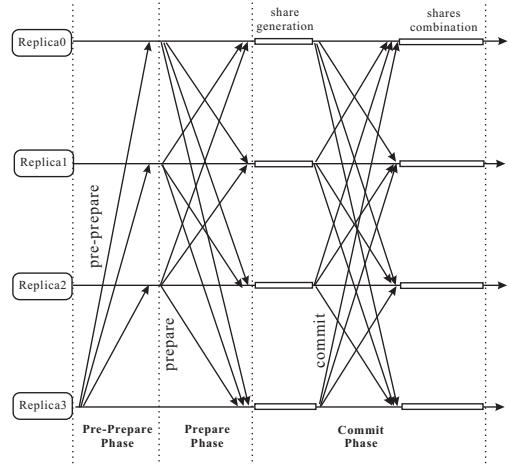}
   \caption{Normal operation of the CT-algorithm.}
   \label{ctfig}
\end{figure}

\section{The CT-Algorithm}
\label{ctsec}

The normal operation of the CT-algorithm is shown in Figure~\ref{ctfig}.
The CT-algorithm is the same as the BFT algorithm in the first two phases
(\ie pre-prepare and prepare phases). The commit phase
is modified by incorporating threshold coin-tossing operations.
Most existing $(k,l)$ threshold signature 
schemes~\cite{threshsig.b,threshcrypto,threshsig.f,threshsig.g,shoup} 
can be used for the
CT-algorithm, where $k$ is the threshold number of signature shares
needed to produce the group signature, and $l=3f+1$ is the total 
number of players (\ie replicas in our case) participating the threshold 
signing. In most $(k,l)$ threshold signature schemes, a correct group 
signature can be derived by combining shares from $k=t+1$ players, 
where $t=f$ is the maximum number of corrupted players tolerated. Some 
schemes, such as the RSA-based scheme in~\cite{shoup}, allow the 
flexibility of using up to $k=l-t$ as the minimum number
of shares required to produce the group signature. Since $l=3f+1$ in our
work, $k$ can be set as high as $2f+1$. This property offers additional 
protection against Byzantine faulty replicas~\cite{shoup}. 

At the beginning of the commit phase, each replica generates its share
of threshold signature by signing $<$$d||n$$>$ using its private key share,
where $d$ is the digest of the request message and $n$ is the sequence
number assigned to the request. This operation is referred to as the
share-generation step in Figure~\ref{ctfig}. The signature share is
piggybacked with the commit message, in the form 
$<${\sc commit}$v,n,i,d,T(d||n,i)$$>$$\alpha_{i}$, where $T(d||n,i)$ is the
replica $i$'s share of threshold signature.

When a replica has collected $2f$$+$$1$ valid commit messages from different
replicas, it executes the shares-combination step by combining $k$ 
threshold signature shares piggybacked with the commit messages.
After the shares have been combined into a group signature, it is mapped 
into a random number, first by hashing the group signature with a secure
hash function (\eg SHA1), and then by 
taking the first group of most significant bits from the
hash according to the type of numbers needed, \eg 32bits. The
random number will be delivered together with the request to the application,
when all previous requests have been delivered.

\section{Informal Proof of Correctness}
In this section, we provide an informal argument on the correctness
of our two algorithms. The correctness criteria for the algorithms
are:
\begin{itemize}
\item[C1]All correct replicas deliver the same random number to the 
application together with the associated request, and
\item[C2]The random number is secure (\ie it is truly random) in the 
presence of up to $f$ Byzantine faulty replicas.
\end{itemize}

We first argue for the BA-algorithm. C1 is guaranteed by the use of
Byzantine agreement algorithm. C2 is ensured by the collection of 
$2f+1$ shares
contributed by different replicas, and by a sound entropy combination
algorithm (\eg by using the bitwise exclusive-or operation on the set
to produce the combined random number). By collecting $2f+1$ 
contributions, it is guaranteed that at least $f+1$ of them are
from correct replicas, so faulty replicas cannot completely
control the set.\footnote{The use of $f+1$ shares are all that needed
for this purpose. However, collecting more shares is more robust in
cases when some correct replicas use low-entropy sources. This is analogous
to the benefit of Shoup's threshold signature scheme~\cite{shoup}.}
The entropy combination algorithm ensures that the combined random
number is secure as long as at least one share is secure. The
bitwise exclusive-or operation could be used to combine the set
and it is provably secure for this purpose~\cite{cryptovirology}. 
Therefore, the BA-algorithm satisfies both C1 and C2.

Next we argue for the CT-algorithm. C1 is guaranteed by the following
fact: (1) The same message ($<$$d||n$$>$) is signed by all correct replicas, 
according to the CT-algorithm. (2) The threshold signature algorithm 
guarantees the production of the same group signature by combining 
$k$ shares. Different replicas could obtain different set of $k$ shares
and yet they all lead to the same group signature.
(3) The same secure hash function is used to hash the group signature.
C2 is guaranteed by the threshold signature algorithm. For the 
threshold signature algorithm used in our implementation, its security is 
ensured by the random oracle model~\cite{shoup}. Therefore, the 
CT-algorithm is correct as well. This completes our proof.

\section{Performance Characterization}
\label{implsec}
The BA-algorithm and the CT-algorithm have been implemented and incorporated
into a Java-based BFT framework. The Java-based BFT framework is developed
in house and it is ported from the C++ based BFT framework 
of Castro and Liskov~\cite{bft-acm}. Due to space limitation, the details
of the framework implementation is omitted. 
The CT-algorithm uses Shoup's threshold
signature scheme~\cite{shoup}, implemented by Steve Weis and made 
available at SourceForge~\cite{ctcode}.

The development and test platform consists of a group of Dell SC440 servers
each is equipped with a PentiumD processor of 2.8GHz and 1GB of RAM running 
SuSE 10.2 Linux. The nodes are connected via a 100Mbps LAN.
As we noted earlier, the WAN experiments are emulated by introducing
artificial delays in communication, without injecting message loss.

To character the cost of the two algorithms, we use an 
echo application with fixed 1KB-long requests and replies. The server is 
replicated at four nodes, and hence, $f=1$ in all our measurements. Up to 12 
concurrent clients are launched across the remaining nodes (at most one 
client per node). Each client issues consecutive requests without any
think time.
For the CT-algorithm, we vary a number of parameters, including the threshold
value and the key length. We also experiment with certain optimizations.
For all measurements, the end-to-end latency is measured at the client
and the throughput is measured at the replicas. The Java 
{\tt System.nanoTime()} API is used for all timing-related measurements.

\subsection{Cost of Cryptographic Operations}
We first report the mean execution latency of basic cryptographic operations 
involved in the BA-algorithm and the CT-algorithm because
such information is beneficial to the understanding of the behaviors
we observe. The latency cost is obtained when running a single client
and 4 server replicas in the LAN testbed. The results are summarized in
Table~\ref{cryptocost}. As can be seen, the threshold signature
operations are quite expensive, and it is impractical to use a key as 
large as 1024bit-long. 

\begin{table}[t]
\begin{center}
  \begin{tabular}{|| p{0.7in} | p{0.7in} | p{0.7in} ||}
    \hline
\small    Operation Type & Signing /Generation
& \small Verification /Combination \\ \hline\hline
\small    MAC& \small {24.1 ${\mu}s$} &\small 237.3 ${\mu}s$\\ \hline
\small    Authenticator& \small 80.2${\mu}s$ &\small 892.0${\mu}s$\\ \hline
\small    CT2-64 & \small 2.2$ms$ &\small 4.6$ms$\\ \hline
\small    CT2-128 & \small 7.1$ms$ &\small 12.8$ms$\\ \hline
\small    CT2-256 & \small 31.7$ms$ &\small 58.5$ms$\\ \hline
\small    CT2-512 & \small 179.1$ms$ &\small 338.2$ms$\\ \hline
\small    CT2-1024 & \small 1191.7$ms$ &\small 1381.4$ms$\\ \hline
\small    CT3-64 & \small 2.2$ms$ &\small 5.6$ms$\\ \hline
\small    CT3-128 & \small 7.1$ms$ &\small 18.5$ms$\\ \hline
\small    CT3-256 & \small 31.7$ms$ &\small 80.0$ms$\\ \hline
\small    CT3-512 & \small 179.1$ms$ &\small 449.7$ms$\\ \hline
\small    CT3-1024 & \small 1191.7$ms$ &\small 2292.1$ms$\\ \hline
    \hline
  \end{tabular}
\caption{Execution time for basic cryptographic operations involved
with our algorithms. The data shown for CT signing is for a single share.}
\label{cryptocost}
\end{center}
\end{table}

Without any optimization (and without fault), an end-to-end remote call
from a client to the replicated server using the original BFT algorithm
involves a total of 4 authenticator generation operations ($A_g$), 
5 authenticator verification operations ($A_v$) (one does not need to
verify the message sent by itself), 1 MAC generation
operation ($M_g$) and 2 MAC verification operation ($M_v$) on the critical
execution path (\ie $A_g+A_v$ for
request sending and receiving, $A_g+A_v$ for the pre-prepare phase,
$A_g+A_v$ for the prepare phase, $A_g+2A_v$ for the commit phase, and
$M_g+2M_v$ for the reply sending and receiving). The BA-algorithm introduces 
two additional communication steps and 2 $A_g$ and 3 $A_v$ on the critical 
path. The CT-algorithm does not require any additional communication step, 
but introduces 1 threshold signing operation ($T_s$) and 1 operation for 
threshold shares verification and 
combination ($T_v$). From this analysis, the minimum end-to-end latency 
achievable using the BA-algorithm is $L_{BA}^{min}=6A_g+8A_v+M_g+2M_v$
(a replica can proceed to the next step as soon as it receives 1 valid
prepare message from other replica in the prepare phase, and 2 valid
commit messages from other replicas in the commit phase, and the client
can proceed to deliver the reply as soon as it has gotten 2 consistent
replies). Similarly, the minimum latency using the CT-algorithm is 
$L_{CT}^{min}=4A_g+5A_v+M_g+2M_v+T_s+T_v$. Based on the values given in
Table~\ref{cryptocost}, $L_{BA}^{min}=8.1ms$ and $L_{CT}^{min}=12.1ms$ for
$k=2$ and 64bit-long key. The minimum overhead incurred by the BA-algorithm
is $2A_g+3A_v=2.8ms$ and that by the CT-algorithm is $T_s+T_v=6.8ms$ for
$k=2$ and 64bit-long key.

\begin{figure}[t]
  \center
   \includegraphics[width=2.6in]{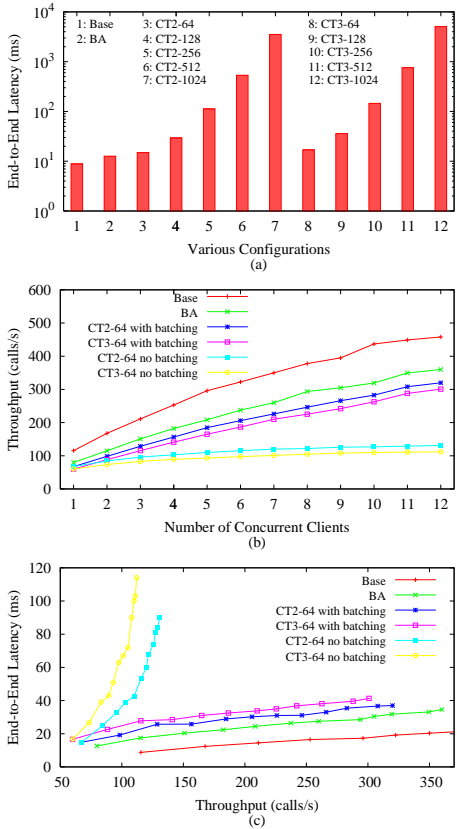}
   \caption{LAN measurement results. (a) End-to-end latency
under various configurations. (b) The system throughput in the presence
of different number of concurrent clients. (c) End-to-end latency as
a function of the load on the system (throughput).}
   \label{lanperf}
\vspace{-0.3in}
\end{figure}

\subsection{LAN Experimental Results}

Figure~\ref{lanperf} shows the summary of the experimental results
obtained in the LAN testbed. The end-to-end latency (plotted in log-scale)
measured at a single
client under various configurations is shown in Figure~\ref{lanperf}(a). 
As a reference, the latency for the BFT system without the additional
mechanisms described in this paper is shown as ``Base''.
In the figure, the result for the BA-algorithm is shown as ``BA'',
and the results for the CT-algorithm with different parameter settings
are labeled as CT\#-i, where \# is the $k$ value, and $i$ is the key length.
As can be seen, only if a very short key is used, the CT-algorithm incurs
significant overhead. Furthermore, the observed
end-to-end latency results are in-line with the analysis provided
in the previous subsection. 

The throughput measurement results shown in Figure~\ref{lanperf}(b) 
are consistent with those in the end-to-end latency measurements. 
The results labeled with ``no batching'' are
obtained for the original CT-algorithm described in Section~\ref{ctsec},
\ie one coin-tossing operation (\ie threshold share signing, combination
and verification of $k$ shares) is used for {\em every} request. Those
labeled with ``with batching'' are measured when the requests are batched
(for total ordering, they all share the same sequence number~\cite{bft-acm})
and only one coin-tossing operation is used for the entire batch of requests.
As can be seen from Figure~\ref{lanperf}(b), the gain in throughput is 
significant with the batching optimization. However, if sharing the same
random number among several requests is a concern, this optimization must
be disabled. 

For the BA-algorithm, the communication steps for reaching a Byzantine 
agreement on the set of random numbers are automatically batched together 
with that for requests total-ordering. Batching the Byzantine agreement 
for a set of random numbers 
does not seem to introduce any vulnerability. The additional optimization
of one set of entropy extraction and combination per batch of requests does 
not have any noticeable performance benefit. Therefore, it is advised that
this further optimization not to be considered in practice due to possible 
security concerns.

Figure~\ref{lanperf}(c) shows the end-to-end latency as a function of the
load on the system in the presence of concurrent clients. We use the system
throughput as a metric for the system load because it better reflects
the actual load on the system than the number of clients. It is also useful
to compare with the results in the WAN experiments. As can be seen,
for the CT-algorithm, without the batching optimization, the latency 
increases very sharply with
the load, due to the CPU intensive threshold signature computations.

The results for the CT-algorithm with keys larger than 64bits are
omitted in Figure~\ref{lanperf}(b) and (c) to avoid cluttering. 
The throughput is significantly lower and the end-to-end latency is
much higher than those of the BA-algorithm in these configurations, 
especially when the load is high.

\begin{figure}[t]
  \center
   \includegraphics[width=2.6in]{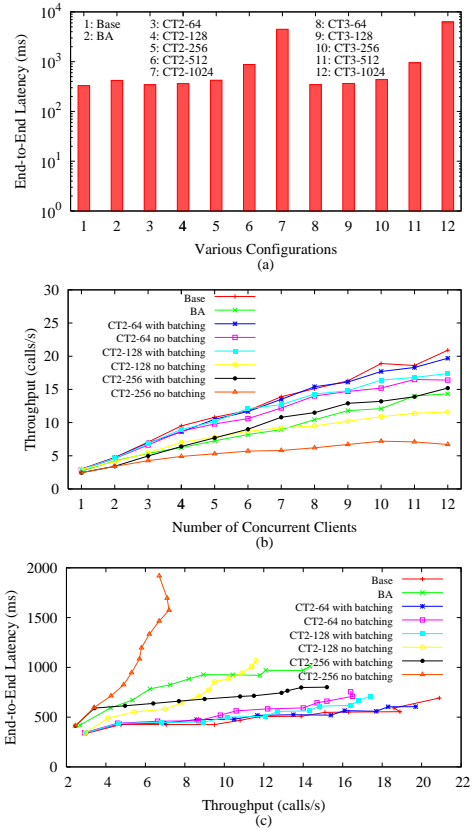}
   \caption{Emulated WAN measurement results. (a) End-to-end latency
under various configurations. (b) The system throughput in the presence
of different number of concurrent clients. (c) End-to-end latency as
a function of the load on the system (throughput).}
   \label{wanperf}
\vspace{-0.1in}
\end{figure}

\subsection{WAN Experimental Results}

The experimental results obtained in an emulated WAN environment are shown
in Figure~\ref{wanperf}. The observed metrics and the parameters used are
identical to those in the LAN experiments. As can be seen in 
Figure~\ref{wanperf}(a), the end-to-end latency as perceived by a single 
client is similar for the BA-algorithm and the CT-algorithm with a key size
up to 256bits (for either $k=2$ or $k=3$). This can be easily understood
because the end-to-end latency is dominated by the communication delays,
as indicated by the end-to-end latency for the base system included in
the figure. 

Figure~\ref{wanperf}(b) shows part of the measurement results on system
throughput under different number of concurrent clients. To avoid 
cluttering, only the results for $k=2$ and key sizes of up to 256bits
are shown . The throughput for the base system is included as a reference.
As can be seen, when batching for the coin-tossing operation is enabled,
the CT-algorithm with short-to-medium sized keys out-performs the 
BA-algorithm. When batching is disabled, however, the BA-algorithm
performs better unless very small key is used for the CT-algorithm.
The end-to-end latency results shown in Figure~\ref{wanperf}(c) confirm
the trend.

\section{Related Work}
\label{rwsec}

How to ensure strong replica consistency in the presence of replica
nondeterminism has been of research interest for a long time, especially
for fault tolerant systems using the benign fault 
model~\cite{bft-acm, base,delta4,
priya06}. However, while the importance of the use of good random numbers
has long been recognized in building secure systems~\cite{ssbook}, we have
yet to see substantial research work on how to preserve the randomized
operations necessary to ensure the system integrity in a fault tolerant
system. For the type of systems where the use of random numbers is crucial
to their service integrity, the benign fault model is obviously inadequate
and the Byzantine fault model must be employed if fault tolerance is 
required.

In the recent several years, significant progress has been made towards
building practical Byzantine fault tolerant systems, as shown in the
series of seminal papers such 
as~\cite{bft-acm,base,speculativebft,alvisi-bft}. This makes it possible
to address the problem of reconciliation of the requirement
of strong replica consistency and the preservation of each replica's 
randomness for real-world applications that requires both high 
availability and high degree of security. We believe the work presented 
in this paper is an important step towards solving this challenging problem.

We should note that some form of replica nondeterminism (in particular,
replica nondeterminism related to timestamp operations) has been studied
in the context Byzantine fault tolerant systems~\cite{bft-acm, base}.
However, we have argued in previous sections that the existing approach
is vulnerable to the presence of colluding Byzantine faulty replicas and
clients. 

The main idea of this work, \ie collective determination of random
values based on the contributions made by the replicas, is borrowed
from the design principles for secure communication protocols~\cite{cnbook}. 
However, the application of this principle in solving the strong replica
consistency problem is novel.

The CT-algorithm is inspired by the work of Cachin, Kursawe and 
Shoup~\cite{ctba}, in particular, the idea of exploiting 
threshold signature techniques for agreement. 
However, we have adapted this idea to solve a totally different problem,
\ie it is used towards reaching
integrity-preserving strong replica consistency. Furthermore,
we carefully studied what to sign for each request so that the final 
random number obtained is not vulnerable to attacks.

\section{Conclusion and Future Work}
\label{concsec}
In this paper, we presented our work on reconciling the requirement of
strong replica consistency and the desire of maintaining each replica's
individual randomness. Based on the central idea of collective
determination of random values needed by the applications for their
service integrity, we designed and implemented two algorithms. The first
one, the BA-algorithm, is based on reaching a Byzantine agreement on a set
of random number shares provided by $2f+1$ replicas. The second one,
the CT-algorithm, is based on threshold signature techniques.
We thoroughly characterized the performance of the two algorithms in
both a LAN testbed and an emulated WAN environment. We show that the
BA-algorithm in general out-performs the CT-algorithm in most cases
except in WAN operations under relatively light load. Furthermore,
the overhead incurred by the BA-algorithm with respect to the base
BFT system is relatively small, making it possible for practical use.

Future research work will focus on the threshold key share refreshment
issue for the CT-algorithm. To ensure long-term robustness of the
system, the key shares must be proactively refreshed periodically.
Otherwise, the random numbers generated this way may age over time,
which may open the door for attacks. The threshold signature
algorithm used in this work~\cite{shoup} does
not have built-in mechanism for key share refreshment. We will explore
other threshold signature algorithms that offer this capability~\cite{threshsig.b,threshsig.f,threshsig.g}. 



\begin{thebibliography}{10}\setlength{\itemsep}{-1ex}\small

\bibitem{threshsig.b}
A. Boldyreva.
\newblock Efficient threshold signatures, multisignatures and blind 
signatures based on the Gap-Diffie-Hellman-Group signature scheme.
\newblock {\em Lecture Notes in Computer Science}, 2567:31--46, Springer-Verlag, 
2003

\bibitem{ctba}
C.~Cachin, K.~Kursawe, and V.~Shoup.
\newblock Random oracles in Constantinople:
Practical asynchronous Byzantine agreement using cryptography.
\newblock {\em Journal of Cryptology}, 18:219--246, 2005.

\bibitem{bft-acm}
M.~Castro and B.~Liskov.
\newblock Practical {Byzantine} fault tolerance and proactive recovery.
\newblock {\em ACM Transactions on Computer Systems}, 20(4):398--461, November
  2002.

\bibitem{base}
M.~Castro, R.~Rodrigues, and B.~Liskov.
\newblock {BASE}: Using abstraction to improve fault tolerance.
\newblock {\em ACM Transactions on Computer Systems}, 21(3):236--269, August
  2003.

\bibitem{threshcrypto}
Y. Desmedt.
\newblock Threshold cryptography.
\newblock {\em European Transactions on Telecommunications}, 5(4):449--457
1994.

\bibitem{deswarte}
Y. Deswarte, L. Blain, and J.-C. Fabre.
\newblock Intrusion tolerance in distributed computing systems.
\newblock {\em Proceedings of the IEEE Symposium on Research in Security 
and Privacy}, pages 110--121, Oakland, CA, May 1991.

\bibitem{threshsig.f}
Y. Frankel, P. Gemmal, P. MacKenzie and M. Yung.
\newblock Proactive RSA.
\newblock {\em Proceedings of the 17th Annual International Cryptology 
Conference (Crypto' 97)}, Santa Barbara, CA, August 1997.

\bibitem{threshsig.g}
R. Gennaro, S. Jarecki, H. Krawczyk and T. Rabin.
\newblock Robust threshold DSS signatures.
\newblock {\em Proceedings of the International Conference on the Theory 
and Application of Cryptographic Techniques}, Saragossa, Spain, May 12-16, 
1996.

\bibitem{speculativebft}
R. Kotla, L. Alvisi, M. Dahlin, A. Clement, E. Wong.
\newblock Zyzzyva: Speculative Byzantine fault tolerance.
\newblock In {\em Proceedings of 21st ACM Symposium on Operating Systems 
Principles}, WA, 2007.

\bibitem{truerand}
J. Lacy, D. Mitchell, and W. Schell.
\newblock CryptoLib: Cryptography in software.
\newblock {\em Proceedings of the 4th USENIX Security Symposium}, 
pages 1--17, 1993.

\bibitem{lamport:byz}
L.~Lamport, R.~Shostak, and M.~Pease.
\newblock The {B}yzantine generals problem.
\newblock {\em ACM Transactions on Programming Languages and Systems},
  4(3):382--401, July 1982.

\bibitem{delta4}
D.~Powell.
\newblock {\em {Delta-4}: A Generic Architecture for Dependable Distributed
  Computing}.
\newblock Springer-Verlag, 1991.

\bibitem{threshsig.r}
T. Rabin, 
\newblock A simplified approach to threshold and proactive RSA.
\newblock {\em Proceedings of the 18th Annual International Cryptology 
Conference}(Crypto' 98), Santa Barbara, CA, August 1998.

\bibitem{shoup}
V. Shoup.
\newblock Practical threshold signatures.
\newblock {\em Lecture Notes in Computer Science},
1097:207--220, Springer, Berlin, 2000.

\bibitem{priya06}
J.~Slember and P.~Narasimhan.
\newblock Living with nondeterminism in replicated middleware applications.
\newblock In {\em Proceedings of the ACM/IFIP/USENIX 7th International
  Middleware Conference}, pages 81--100, Melbourne, Australia, 2006.

\bibitem{statemachine}
F. Schneider.
\newblock Implementing fault-tolerant services using the state machine 
approach: a tutorial.
\newblock {\em ACM Computing Surveys}, 22(4):299--319, 1990.

\bibitem{cnbook}
A. Tanenbaum.
\newblock {\em Computer Networks}, Prentice Hall, 2003, 4th Edition.

\bibitem{ctcode}
ThreshSig: Java threshold signatures.
Available at http://threshsig.sourceforge.net/

\bibitem{ssbook}
J.~Viega and G.~McGraw.
\newblock {\em Building Secure Software}.
\newblock Addison-Wesley, 2002.

\bibitem{alvisi-bft}
J.~Yin, J.-P. Martin, A.~Venkataramani, L.~Alvisi, and M.~Dahlin.
\newblock Separating agreement from execution for byzantine fault tolerant
  services.
\newblock In {\em Proceedings of the ACM Symposium on Operating Systems
  Principles}, pages 253--267, Bolton Landing, NY, USA, 2003.

\bibitem{cryptovirology}
A. Young and M. Yung.
\newblock {\em Malicious Cryptography: Exposing Cryptovirology}.
\newblock Wiley, 2004.

\end{thebibliography}
\end{document}